\documentclass{elsart1p}

\usepackage{graphics}
\usepackage{graphicx}
\usepackage{epsfig}
\usepackage{amssymb}

\newcommand{\be}{\begin{equation}}
\newcommand{\ee}{\end{equation}}
\newcommand{\ba}{\begin{eqnarray}}
\newcommand{\ea}{\end{eqnarray}}
\newcommand{\nn}{\nonumber}
\newcommand{\ex}{{\rm e}}
\newcommand{\Tr}  {\mathop{\rm Tr}}

\def\lsi{\raise0.3ex\hbox{$<$\kern-0.75em\raise-1.1ex\hbox{$\sim$}}}
\def\gsi{\raise0.3ex\hbox{$>$\kern-0.75em\raise-1.1ex\hbox{$\sim$}}}

\newcommand{\eq}{Eq.~}
\newcommand{\eqs}{Eqs.~}
\newcommand{\fig}{Fig.~}
\def\bfx{{\bf x}}
\def\bfy{{\bf y}}
\def\bfr{{\bf r}}
\newcommand{\Nc}{N_{\rm c}}
\newcommand{\re}{\mathop{\mbox{Re}}}
\newcommand{\im}{\mathop{\mbox{Im}}}

\begin{document}

\begin{flushright}
MS-TP-08-27
\end{flushright}

\begin{frontmatter}

% Title, authors and addresses

% use the thanksref command within \title, \author or \address for footnotes;
% use the corauthref command within \author for corresponding author footnotes;
% use the ead command for the email address,
% and the form \ead[url] for the home page:
% \title{Title\thanksref{label1}}
% \thanks[label1]{}
% \author{Name\corauthref{cor1}\thanksref{label2}}
% \ead{email address}
% \ead[url]{home page}
% \thanks[label2]{}
% \corauth[cor1]{}
% \address{Address\thanksref{label3}}
% \thanks[label3]{}

\title{Static potentials for quarkonia at finite temperatures}

% use optional labels to link authors explicitly to addresses:
% \author[label1,label2]{}
% \address[label1]{}
% \address[label2]{}

\author{Owe Philipsen}

\address{Institut f\"ur Theoretische Physik, 
Universit\"at M\"unster, 48149 M\"unster, Germany}

\begin{abstract}
We review non-perturbative static potentials commonly used in potential
models for quarkonia at finite $T$. Potentials
derived from Polyakov loop correlators are shown to be inappropriate for 
this purpose. The $q\bar{q}$ free energy is physical but has the wrong
spatial decay and perturbative limit.
% differ from the known Debye-screened potential. 
The so-called singlet free energy is gauge dependent and  
%, and its difference from the average free energy is 
unphysical. An appropriate static real time potential can be defined through
a generalisation of pNRQCD to finite $T$. In perturbation theory, 
its real part reproduces the Debye-screened
potential, its imaginary part accounts for Landau damping. 
Possibilities for its non-perturbative evaluation are discussed.
%Non-perturbative aspects
%of this new potential are discussed. 
\end{abstract}

\end{frontmatter}

% main text
\section{Introduction}
\label{sec:intro}

The properties of quarkonia are believed to provide a useful probe of 
the QCD plasma at high temperatures, in particular for the quark-hadron transition.
This expectation was originally based
on a potential model \cite{mats}, in which the
linearly confining potential for zero temperature gets replaced by a 
Debye-screened potential at high $T$. 

Potential models have a long history for
the description of quarkonia at zero temperature. 
The basic idea is that for heavy quarks of mass $M$, which move non-relativistically, 
%the bound state dynamics is akin to the hydrogen problem.
%In particular, 
the binding energy $(E-2M)$ is small compared to $M$ and can be obtained by solving a 
static Schr\"odinger equation
\be
\left(\frac{\nabla^2}{M}+V(r)\right)\psi=(E-2M)\psi.
\ee
$V(r)$ is the (radially symmetric)
potential between the static quark anti-quark pair separated by a distance $r$. 
Initially $V(r)$ was modelled by the Cornell potential 
(Coulomb plus linear), more recently non-perturbative
lattice data are used as input. The crucial observation is that
the Schr\"odinger equation follows from an effective theory approach. 
Starting from QCD, one can use of the
scale separation between the heavy quark mass $M$ and the binding energy 
$E-2M$, to obtain an effective theory, pNRQCD \cite{pnrqcd}, 
for the low energy dynamics
in the confining potential. In this framework, the static potential appears
as a perturbative matching coefficient of the effective theory.  
Hence, the Schr\"odinger equation can be improved systematically by computing
higher order terms in the scale hierarchy. 
Note, that a very successful spectroscopy with $\sim 1\%$ accuracy
is obtained in this way.

It is tempting to employ this approach also at finite $T$.
Matsui and Satz heuristically used the same equation, but with
a Debye-screened potential from perturbation theory,
\be
V(r,T)\approx -\frac{g^2C_F}{4\pi}\frac{\ex^{-m_D(T)r}}{r}.
\label{deb}
\ee
However, there are a number of problems. Firstly, it is not clear 
if the bound state Schr\"odinger equation 
can be translated to a finite $T$ many body situation,
in a way that temperature effects show up only in the potential. Secondly,
at finite T there exists a variety of non-perturbative 
potentials, and it is not clear which one constitutes the non-perturbative
generalisation of \eq(\ref{deb}).

\section{Static potentials from the lattice at zero and finite T}

At $T=0$, the static potential can be defined non-perturbatively
on a euclidean $L^3\times N_\tau$ space time lattice. 
Consider a meson correlation function with an
interpolating operator $\bar{\psi}(\bfx)U(\bfx,\bfy)\psi(\bfy)$, where $U$
denotes a straight line gauge string between the quarks.
In the limit $M\rightarrow \infty$ the heavy quarks can be integrated out,
taking the correlator to the euclidean Wilson loop,
\be
\langle
\bar{\psi}(\bfx,\tau)U(\bfx,\bfy;\tau)\psi(\bfy,\tau)
\bar{\psi}(\bfx,0)U(\bfx,\bfy;0)\psi(\bfy,0)
\rangle
\longrightarrow \ex^{-2M\tau} W_E(|\bfx-\bfy|)\;.
\ee
Inserting a complete set of eigenstates of the Kogut-Susskind 
Hamiltonian (in temporal gauge), the Wilson loop evaluates to 
($r=|\bfx-\bfy|, U_r\equiv U(\bfx,\bfy;0)$)
\ba
\label{spec}
W_E(r,\tau)&=&\frac{1}{Z}\sum_{n,m}|\langle n|U_r|m\rangle|^2 
\ex^{-E_nN_\tau}\,\ex^{-(E_m(r)-E_n)\tau}\\ 
\hspace*{-3cm}
&\stackrel{N_\tau\rightarrow \infty}{\longrightarrow}&
\sum_m |\langle 0|U_r|m\rangle |^2\,
\ex^{-(E_m(r)-E_0)\tau} 
\stackrel{\tau\rightarrow \infty}{\longrightarrow}
|\langle 0|U_r|1\rangle|^2\,\ex^{-(E_1(r)-E_0)\tau},
\ea
with $E_m(r)$ eigenvalues in the sector with sources, and $E_n$ in the sector without.
On the lattice, $T=1/(aN_\tau)$, hence $T=0$ implies $N_\tau\rightarrow \infty$ in
the second line.
Taking furthermore the limit
$\tau\rightarrow \infty$, the sum is dominated by the lowest energy state.
The static potential is defined to be the lowest energy of the static
quark anti-quark configuration at a given separation, $V(r)\equiv E_1(r)-E_0$.
Note that the matrix element with the string operator is of no interest here.

The generalisation to finite $T$ is difficult to interpret because of
the finite and short temporal extent, $N_\tau=1/(aT)$. Thus, we
have to deal with the full superposition \eq(\ref{spec}), to which now also the matrix
elements contribute, and the result still depends on $\tau$.

\begin{figure}
\hspace*{-1.3cm}
\includegraphics[width=0.38\textwidth]{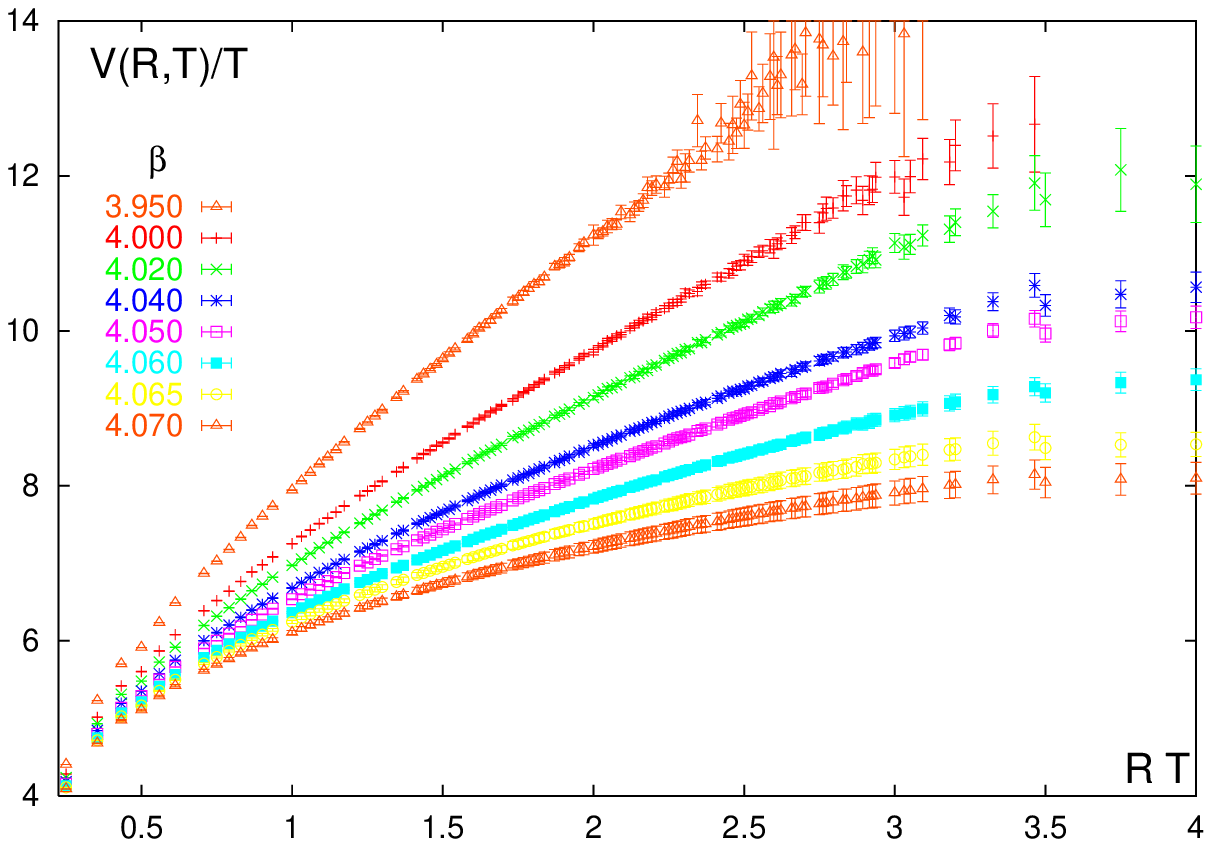}
\includegraphics[width=0.38\textwidth]{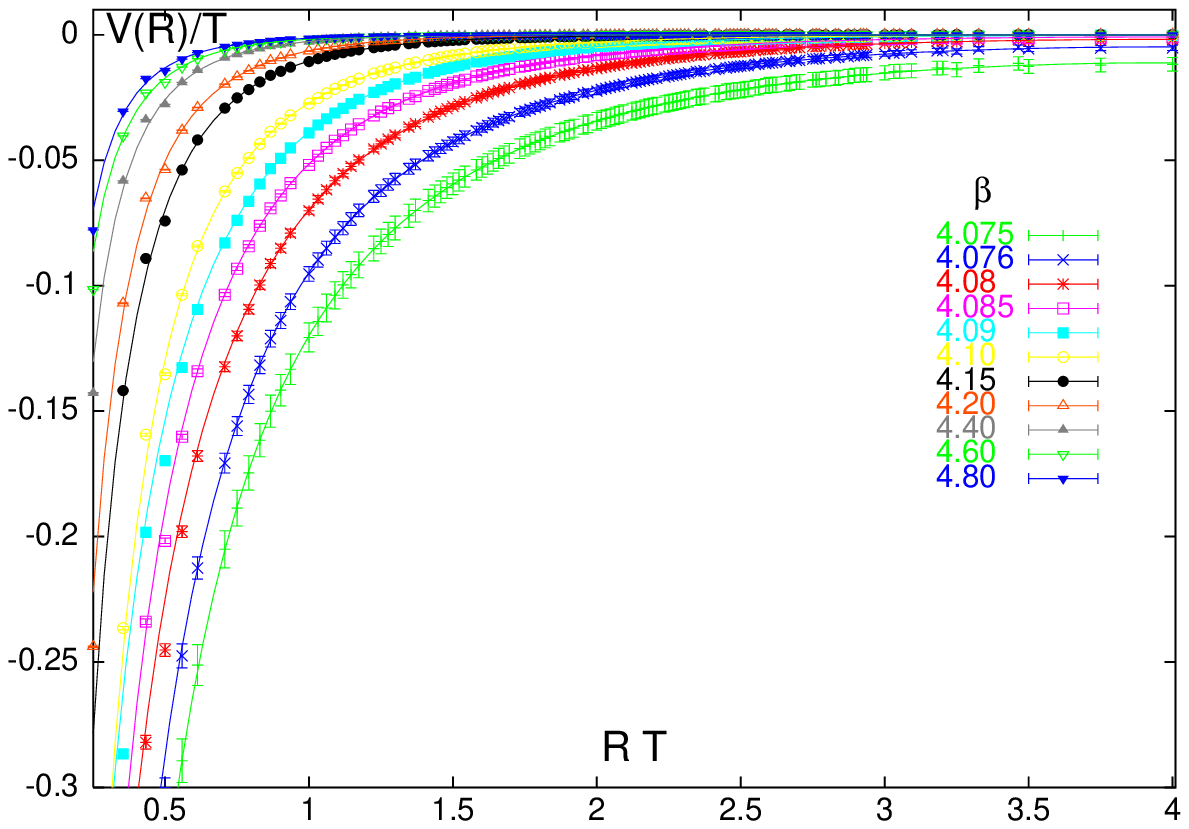}
\includegraphics[width=0.38\textwidth]{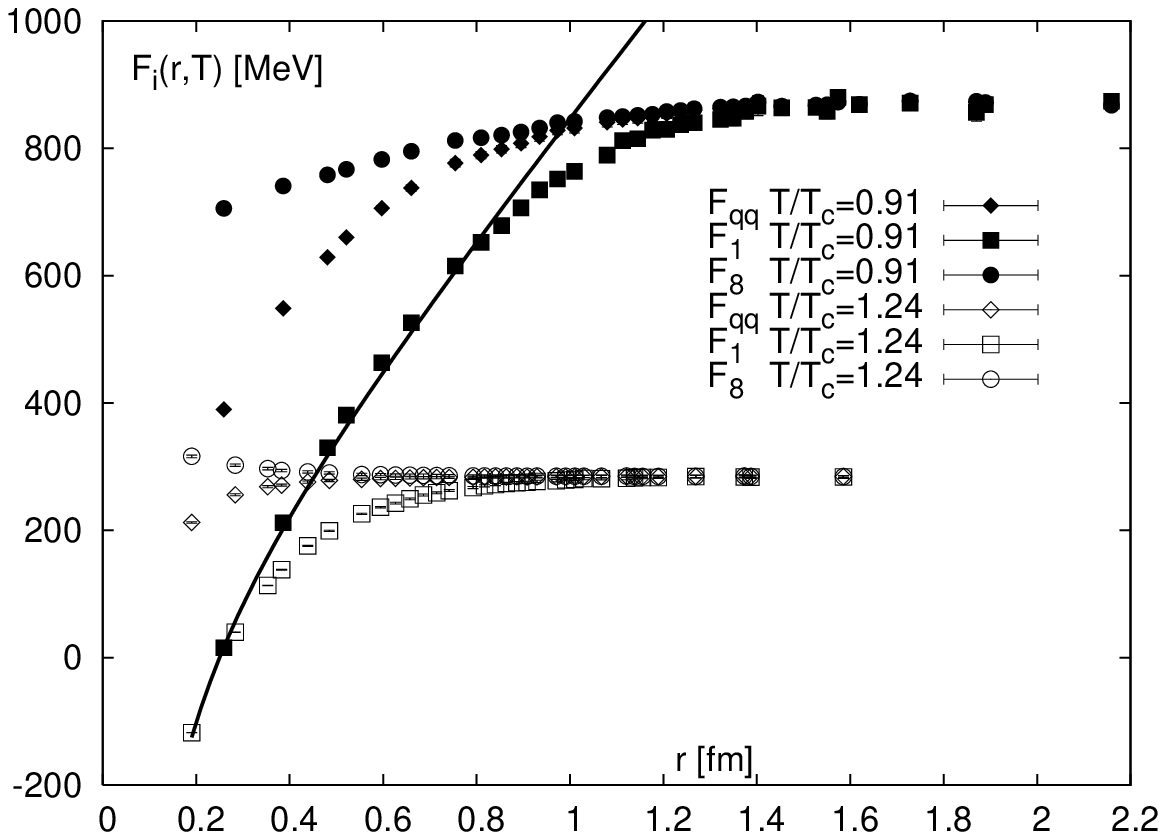}\hspace*{-1.5cm}
\caption[]{Static quark anti-quark free energy/potential, \eq(\ref{av}), for
$T<T_c$ (left) and $T>T_c$ (middle) \cite{potdat}. Right:
Free energies for the three channels \eq(\ref{potdef}). 
The solid line is the 
zero temperature potential \cite{kz}.
}
\label{bie}
\end{figure}

A different definition of the static potential which does generalise to finite $T$
is based on the Polyakov loop 
$L(\bfx)=\prod_{\tau=1,N_\tau} U_0(\bfx,\tau)$, 
i.e.~a static quark sitting at $\bfx$
and propagating in euclidean time through the periodic boundary. 
It transforms in the adjoint, so its trace is gauge invariant.
By spectral analysis one establishes that the Polyakov loop correlator
represents the free energy of a static quark anti-quark
pair separated by $r$ \cite{ls},
\be
\ex^{-F_{\bar{q}q}(r,T)/T}=\frac{1}{N_c^2}\langle \Tr L^\dag (\bfx) \Tr L(\bfy)\rangle
=\frac{1}{ZN_c^2}\sum_{n}\; \ex^{-E_n(r)/T}.
\label{av}
\ee
The energy levels entering this Boltzmann sum are identically the same as
the $E_m(r)$ from the Wilson loop, \eq(\ref{spec}). 
Hence, for $T\rightarrow 0$ we recover $V(r)$, cf.~\cite{lw}.
The free energy is thus often called
a $T$-dependent potential, $V(r,T)\equiv F_{\bar{q}q}(r,T)$. 
The Polyakov loop correlator is readily simulated, with results as in \fig\ref{bie}.
It gives a linear potential in the confined phase, whose string tension
reduces with temperature, while in the deconfined phase the potential is 
screened, \fig\ref{bie}.
Unfortunately, this is {\em not} the Debye-screened potential 
we want, as becomes apparent when considering its spatial decay at high $T$.
Fitting to
\be
\frac{F_{q\bar{q}}}{T}=-\frac{c(T)}{(rT)^d}\,\ex^{-m(T)r},
\ee
gives $d\approx 1.5$ and $m=M_{0^{++}_+}$, i.e.~the screening mass corresponds
to the lightest, gauge-invariant glueball channel \cite{lp}. 
This can already be seen in perturbation theory, 
where the leading term is by two-gluon exchange and thus $m=2m_D$ \cite{nad1}.

It was thus suggested to decompose the Polyakov loop correlator
into channels with relative colour singlet and octet orientations of the 
quark anti-quark pair \cite{ls},
\ba
e^{-F_{q \bar q}(r,T)/T }& =&
\frac{1}{9} \;  e^{- F_{1}(r,T)/T } +
\frac{8}{9}   \; e^{- F_{8}(r,T)/T },\\
\ex^{-F_1(r,T)/T}&=&
\frac{1}{3}\langle \Tr L^\dag(\bfx)L(\bfy)\rangle,\nn\\
\ex^{-F_8(r,T)/T}
&=&\frac{1}{8}\langle \Tr L^\dag(\bfx) \Tr L(\bfy)\rangle
-\frac{1}{24}\langle \Tr L^\dag(\bfx) L(\bfy)\rangle.
\label{potdef}
\ea
Note that the correlators in the singlet and octet channels are gauge
dependent, and the colour decomposition only holds perturbatively in a fixed
gauge. However, in perturbation theory the singlet channel
indeed displays the expected Debye-screened behaviour, 
$F_1(T,r)\sim \ex^{-m_D(T)r}/4\pi r$. This has motivated lattice simulations of
these correlators in fixed Coulomb gauge, 
with results as in \fig\ref{bie} (right). The three different channels show
different $r$-dependence, and hence lead to different binding energies
when used in Schr\"odinger equations. There is a vast literature employing
$F_1$ or the corresponding internal energy $U_1=F_1+TS_1$, and from the solutions
trying to reconstruct lattice meson correlation functions
to check which fits better \cite{mp}. 

However, both options are unphysical at a non-perturbative level.
To understand this, let us start
from something physical and consider a meson
operator in an octet state, 
$O^a=\bar{\psi}(\bfx)U(\bfx,\bfx_0)T^aU(\bfx_0,\bfy)\psi(\bfy)$, with $x_0$
the meson's center of mass.
In the plasma the colour charge can always be neutralised by a gluon.
In the correlators for the singlet and octet operators, we
integrate out the heavy quarks, replacing them by Wilson lines,
\ba
\langle O(\bfx,\bfy;0)O^{\dag}(\bfx,\bfy;N_\tau)\rangle&\propto&
\langle \Tr L^\dag(\bfx)U(\bfx,\bfy;0)L(\bfy)U^\dag(\bfx,\bfy;N_\tau)\rangle,\nn\\
\langle O^a(\bfx,\bfy;0)O^{a\dag}(\bfx,\bfy;N_\tau)\rangle&\propto&
\left[\frac{1}{N_c^2-1}\langle \Tr L^\dag(\bfx) \Tr L(\bfy)\rangle\right.\\
& & \left.
-\frac{1}{N_c(N_c^2-1)}\langle \Tr L^\dag(\bfx)U(\bfx,\bfy;0)
L(\bfy)U^\dag(\bfx,\bfy;N_\tau)\rangle\right]\, . \nn
\ea
We have now arrived at gauge invariant expressions, because we used
a gauge string between the sources. The singlet correlator corresponds
to a periodic Wilson loop which wraps around the boundary. The connection
to the gauge fixed correlators is readily established, replacing 
the gauge string by gauge fixing functions, $U(\bfx,\bfy)=g^{-1}(\bfx)g(\bfy)$. 
Thus, in axial gauge $U(\bfx,\bfy)=1$ (and only there) 
the gauge fixed correlators are identical
to the gauge invariant ones.

\begin{figure}
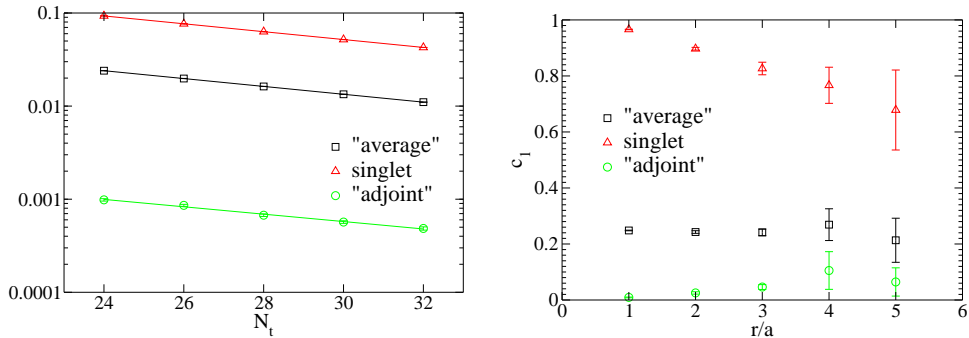

\includegraphics[width=0.45\textwidth]{corrs_r1.eps}\hspace*{0.5cm}
\includegraphics[width=0.45\textwidth]{mat.eps}
\caption[]{\label{3d} Left: Polyakov loop correlators, \eq(\ref{potdef}), for $r/a=1$
in the case of 3d SU(2) in the low temperature limit. All decay 
with the same ground state exponential. Right: The corresponding matrix elements
in the three channels introduces operator dependent $r$-dependence, except for the 
average channel \cite{jp}. 
}
\end{figure}

Next, let us perform the 
spectral analysis. While indeed the energy eigenvalues in the spectral sum 
are independent of the operators \cite{op}, 
the full correlators take the form \cite{jp}
\ba
\ex^{-F_1(r,T)/T}&=&\frac{1}{ZN_c^2}\sum_n\langle n_{\delta\gamma}|
U_{\gamma\delta}(\bfx,\bfy)U^\dag_{\alpha\beta}(\bfx,\bfy)|n_{\beta\alpha}\rangle
\,\ex^{-E_n(r)/T}\;,\nn\\
\ex^{-F_8(r,T)/T}&=&\frac{1}{ZN_c^2}\sum_n\langle n_{\delta\gamma}|
U^a_{\gamma\delta}(\bfx,\bfy)U^{\dag a}_{\alpha\beta}(\bfx,\bfy)|n_{\beta\alpha}\rangle
\,\ex^{-E_n(r)/T}\;.
\label{chan}
\ea
The energy levels in the exponents are identically the same in \eqs(\ref{av},\ref{chan})
and correspond to the familiar gauge invariant static
potential at zero temperature and its excitations. However, while \eq(\ref{av})
is purely a sum of exponentials and thus a true free energy, the 
singlet and octet correlators contain matrix elements which {\em do} depend 
on the operators used, thus giving a path/gauge dependent weight 
to the exponentials contributing to $F_1,F_8$.
This is illustrated numerically in \fig\ref{3d} in the low temperature limit,
where the ground state potential dominates and one can cleanly separate
the exponential and the matrix elements. The $r$-dependent structure is entirely
in the matrix elements, which depend on operators and/or the gauge. 

I do not see how this is evaded by applying smearing techniques,
as recently suggested in \cite{bpv}. These authors 
replace the spatial string swith a smeared object in order to increase the overlap
with the ground state, i.e.~to get the ground state matrix element close to one.
However, most smearing changes the expectation values of correlators, thus destroying
their mutual relations, \eq(\ref{potdef}). Secondly, smearing
increases the weight of the lower energy states at the cost of the
higher ones, and thus undoes the effect of finite temperature in a procedure
dependent way. Finally, if one could get all matrix elements equal to one, 
the different channels would simply be equal, 
up to the trivial colour coefficients, with no additional information.

To summarise, since the spectral information contained in the average and gauge fixed
singlet and octet
channels is the same, we must conclude that any difference between those
correlators is entirely gauge dependent and thus unphysical, and so are all
binding energies calculated in $F_1$ or $U_1$.

\section{A real time static potential for finite T quarkonia}

Progress was made recently by generalising the effective 
theory approach quarkonium physics at $T=0$, namely pNRQCD,
to finite temperatures \cite{us1,bv,soto}, as
reviewed at this conference \cite{laine}.
The analysis is performed in a perturbative setting in Minkowski
time. Just as at zero temperature, the static potential then
appears as a matching coefficient in the effective theory after the heavy
modes have been integrated out. The relevant correlation function is the
quarkonium correlator in real time, but evaluated as a thermal expectation value.
Not surprisingly, after integrating out the static quarks,
the correlator is proportional to a Wilson loop in 
Minkowski time, $W_E(it,\bfr)$.
Of course, the expectation value implied in $W_E$ is now a thermal one,
i.e.~$N_\tau$ is finite for fixed lattice spacing. Hence we need the analytic
continuation of the double spectral sum in \eq(\ref{spec}).
From the effective theory it is easy to see that this correlator obeys
a real time evolution equation
\be
[i\partial_t-V_>(t,r)]W_E(it,r)=0.
\label{schro}
\ee
This represents the desired Schr\"odinger equation for quarkonia
in the plasma, and defines the relevant real time dependent potential. 
The required scale
hierarchy for this equation to be valid is $g^2M<T<gM$. Furthermore, for non-relativistic
bound states $p\ll E$, hence we need $t\gg r$, i.e.~the 
static pontential is obtained in the long time limit $V(\infty,r)$.

\eq(\ref{schro}) may be also be viewed as a non-perturbative definition of the potential
of interest via a correlation function, just as was the case for the zero 
temperature potential. Unfortunately, this one is defined for Minkowski
time and thus requires analytic continuation, i.e.~it cannot be evaluated
directly from euclidean lattice simulations.  
However, a first impression about this object can be gained form HTL-resummed
perturbation theory, for which the leading order result is
\begin{eqnarray}
&&
 V_{>}(\infty,r) = -\frac{g^2 C_F}{4\pi} \biggl[ 
 m_D + \frac{\exp(-m_D r)}{r}
 \biggr] - \frac{i g^2 T C_F}{4\pi} \, \phi(m_D r)\;, \nonumber\\
&&\textrm{with}\hspace{5mm}
 \phi(x) = 
 2 \int_0^\infty \! \frac{{\rm d} z \, z}{(z^2 +1)^2}
 \biggl[
   1 - \frac{\sin(z x)}{zx} 
 \biggr] \;.
\label{pert}
\end{eqnarray}
The most striking feature of this potential is that it is complex, contrary to the free
energies discussed before. The real part features
the expected Debye-screened potential. The imaginary part is due to 
Landau damping and must necessarily be there for a correct effective description of 
the plasma dynamics. Its derivation and properties are discussed in more detail in 
\cite{us1,bv,laine}.

What are the corrections to this potential?
Firstly, there are 
corrections from HTL-resummed perturbation theory
of the order $g^2T/\Lambda$, where $\Lambda$ is the UV cut-off, with a calculable
coefficient. 
Here, we are interested in the non-perturbative corrections from infrared 
modes $\sim g^2T/m_{mag}$. These are due to the soft colour magnetic modes 
$m_{mag}\sim g^2T$, and thus 
cannot be calculated in perturbation theory. 

However, one can calculate
these non-perturbative corrections by classical lattice simulations in that sector
of the theory, which has high occupation numbers and is well represented by a classical
approximation. To identify this sector it is instructive to take the limit 
$\hbar\rightarrow 0$ in our perturbative result \eq(\ref{pert}) first. To this end,
$\hbar$ needs to be reinstated by the replacements $g^2\rightarrow g^2\hbar$,
$1/T\rightarrow \hbar/T$, leading to
\be
\lim_{\hbar\rightarrow 0} V_>(\infty,r)=-\frac{ig^2TC_F}{4\pi}\phi(m_Dr).
\ee
Thus, only the imaginary part survives in the continuum limit. This is easy to understand
since the long range physics of Landau
damping is dominated by classical fields, e.g.~in 
scalar field dynamics, whereas the binding is a generic 
quantum effect, cf.~the hydrogen problem. 
Thus, we can evaluate the non-perturbative infrared effects 
for the imaginary part of the potential.

\section{Imaginary part from classical lattice simulations}

This has been done in \cite{imV}, following the technical setup that was also used
for the evaluation of the sphaleron rate in the electroweak theory \cite{sphal}.
In order to perform real time simulations one reformulates the theory
in a Hamiltonian approach. Fixing temporal gauge $U_0=0$, 
the conjugate field operators are the links and the
electric fields defined by $\dot{U}(\bfx,t)=iE_i(\bfx,t)U_i(\bfx,t)$.
Full gauge invariance is restored by imposing the Gauss constraint
$G(x)\equiv \sum_i
\left[E_i(x)-U_{-i}(x)E_i(x-\hat{i})U_{-i}^\dagger(x)\right]-j^0(x) = 0
$.
A thermal distribution at some initial time is generated by the partition function
\begin{equation}\label{partitionfunction}
 Z=\int \! \mathcal{D}U_i\, \mathcal{D}E_i\, \delta(G) 
 e^{-\beta H}\;,\hspace{5mm}
 H=\frac{1}{\Nc}\sum_x\biggl[ \sum_{i<j} \re\Tr (1-U_{ij})+\frac{1}{2}\Tr( E_i^2 )\biggr]
 \;,
\end{equation}
where $U_{ij}$ is the plaquette. The distribution is then evolved in real time
by the classical equations of motion,
\begin{equation}\label{eom}
 \dot{U}_i(x)=iE_i(x)U_i(x) \;, \quad
 E_i = \sum_a E_i^a T^a \;, \quad
 \dot{E}_i^a(x)=- 2 \im \Tr [T^a\sum_{|j|\neq i} U_{ij}(x)] 
 \;. 
\end{equation}
This procedure can be improved by taking into account quantum corrections 
by using a HTL-resummed effective theory. It will generate a source term
due to the hard particles in the plasma, which modifies the Hamiltonian
as well as the equations of motion by coupling the classical fields
to the quantum effects of the hard particles. That approach was used in 
\cite{sphal} and is easily adapted to the present problem \cite{imV}.

\begin{figure}[t]
  \begin{center}
    \includegraphics[width=0.45\textwidth]{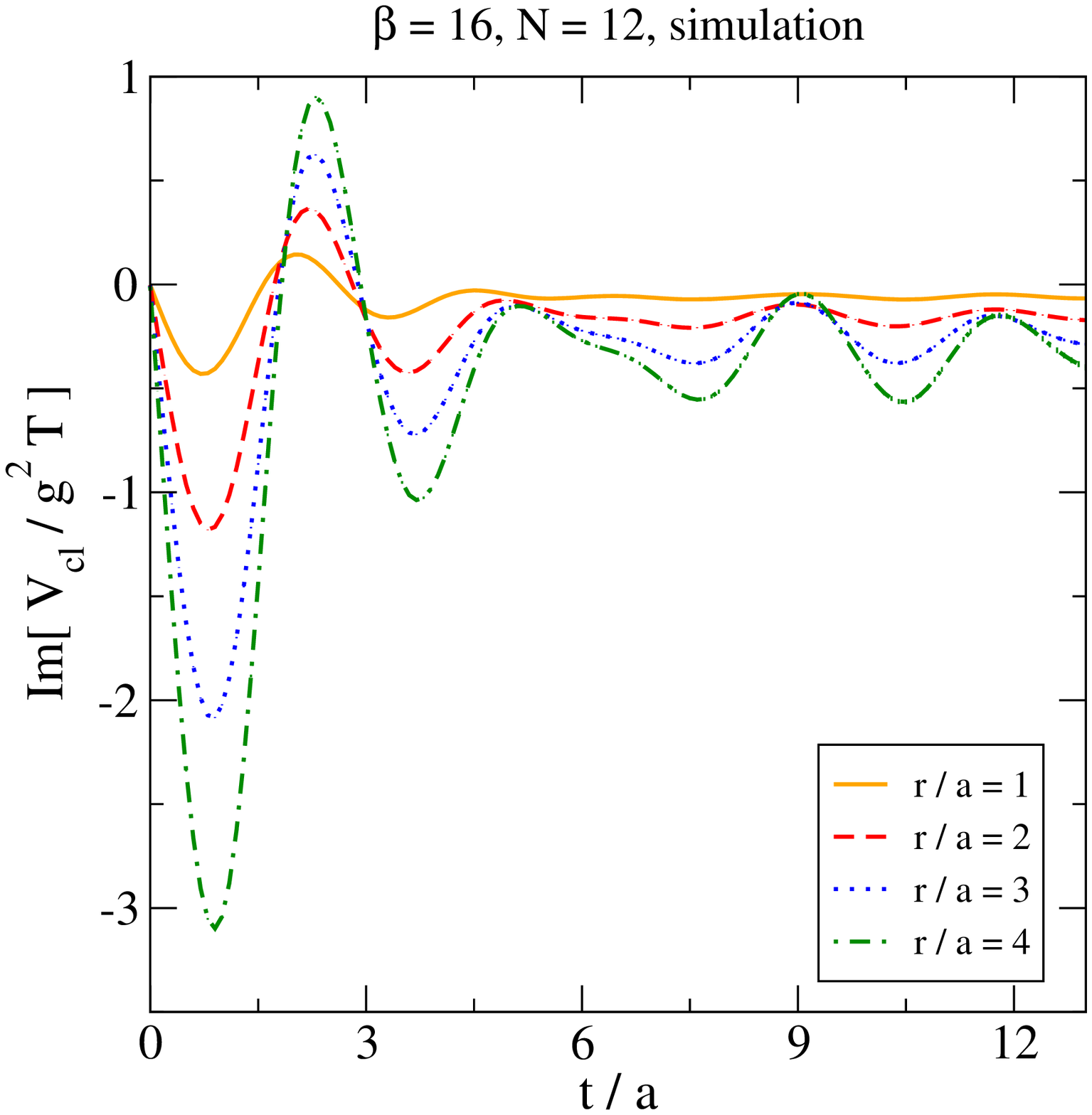}\hspace*{0.5cm}
    \includegraphics[width=0.45\textwidth]{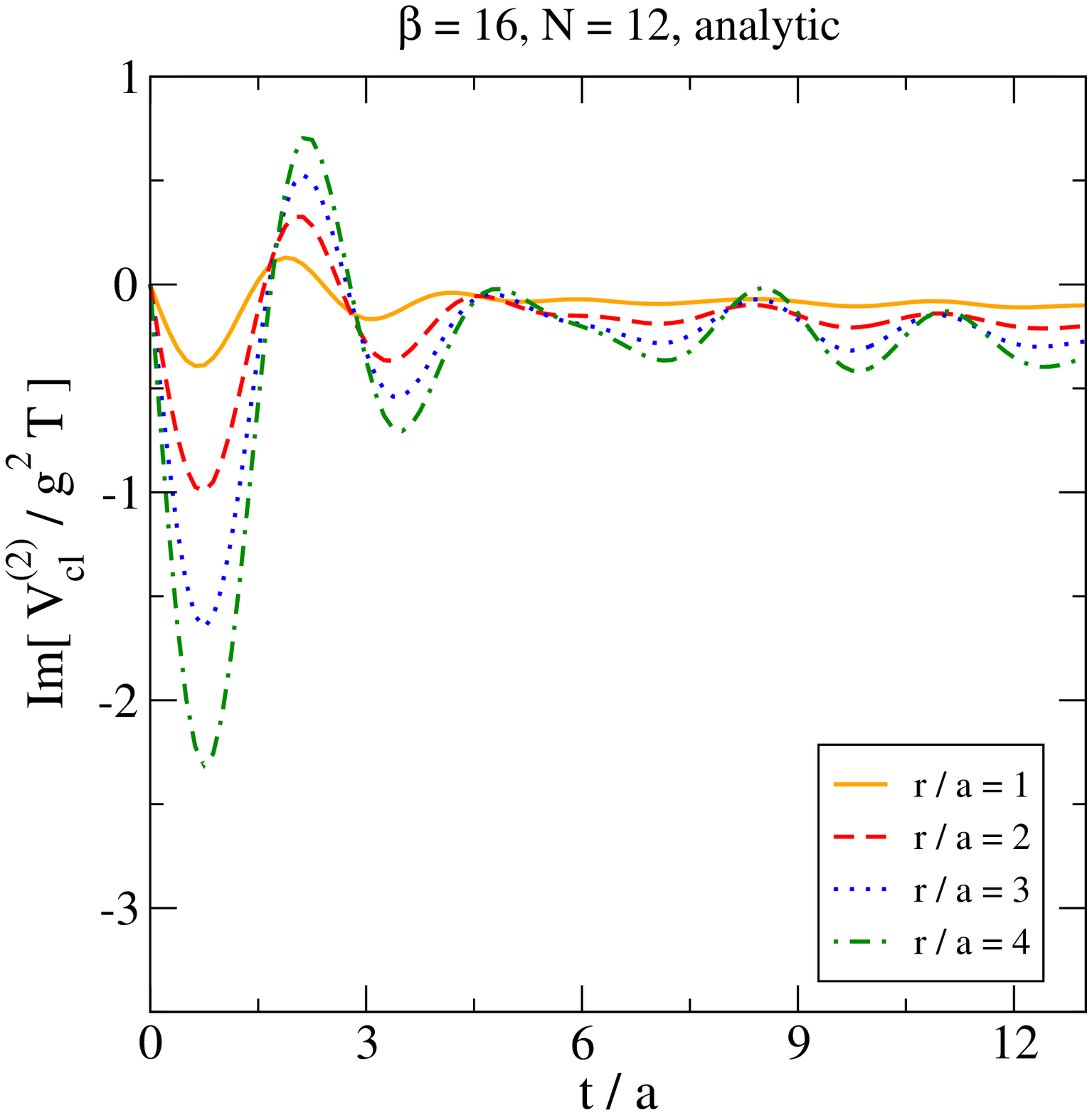}%
  \end{center}\vspace{-4.5mm}
  \caption[]{The imaginary part of the real-time static potential 
from the classical simulation (left) and from resummed perturbation 
theory (right)\cite{imV}.
\label{poti}}
\end{figure}

One can now calculate the classical thermal expectation value of the real time 
Wilson loop, which in temporal gauge reduces to a correlation function of the 
spatial string,
\be
W_{E,cl}(it,r)=\langle U^\dag(\bfx,\bfy;t)U(\bfx,\bfy;0)\rangle.
\ee
The time dependent 
potential extracted via \eq(\ref{schro}) is shown in \fig\ref{poti}, where it
is compared with the result obtained from resummed classical lattice perturbation
theory. We observe complete qualitative agreement. Note in particular that
a finite imaginary part survives also non-perturbatively
in the long time limit, when fluctuations have died
out. Comparing the values one finds that the non-perturbative effects make the 
imaginary part more negative, i.e.~increase Landau damping.

\section{Non-perturbative real time potential?}

It is clearly desirable to go
beyond the classical approximation and construct an operator from which the whole 
quantum potential, including its real part, can be extracted. Of course, a full quantum
computation of the real time Wilson loop is impossible, just as the calculation of
any other real time correlation function. The whole point of the potential approach is 
to bypass the need for such correlators. In our case, what is needed is the static
potential in the infinite time limit, which is clearly less information than
having to know the full time dependence. In terms of correlators, the information we
need is
\be
\lim_{t\rightarrow\infty} W_E(it,r),\quad \lim_{t\rightarrow\infty} \partial_t W_E(it,r).
\ee
At least in principle, 
these limits ought tho be representable by Euclidean
operators, the challenge is to construct those in practice.

\section{Conclusions}

We have argued that many potential models used for the description of 
quarkonia at finite $T$ have significant flaws. 
The connection between the Schr\"odinger equation to 
the underlying quantum field theory is unclear, and lattice
potentials extracted from Polyakov loop correlators, which are typically
used as input for those models, are the wrong quantities for this purpose. 
The average free energy is gauge invariant and well defined, but in its 
perturbative limit does not reduce to the Debye-screened potential. 
The so-called singlet potential is gauge dependent and therefore unphysical. 

These problems can be overcome by using an effective field theory 
obtained by integrating out the heavy quarks, which 
is pNRQCD generalised to finite temperatures. 
The resulting Schr\"odinger equation is the real time
evolution equation for a quarkonium correlator, and the static 
potential in this equation is a matching coefficient in the effective theory.  
To leading order in HTL-resummed perturbation theory, this potential is complex,
its real part showing the correct Debye-screened behaviour and its imaginary
part reflecting Landau damping. 
In the classical limit, only the imaginary part 
survives. This part can be calculated non-perturbatively with
classical lattice simulations in real time. The result agrees in all qualitative features
with the HTL result with slightly strengthened damping. 
It is now important to search for a lattice operator that
represents the real part.\\ 

\noindent
{\bf Acknowledgements:} 
Some of the work presented here is supported by the BMBF project 
\textit{Hot Nuclear Matter from Heavy Ion Collisions
and its Understanding from QCD.}

% The Appendices part is started with the command \appendix;
% appendix sections are then done as normal sections
% \appendix

% \section{}
% \label{}

\end{document}